\pdfoutput=1
\documentclass[12pt]{article}
\usepackage{amsmath}
\usepackage{amssymb}
\usepackage{graphicx} 
\newcommand{\be}{\begin{equation}}
\newcommand{\ee}{\end{equation}}
\newcommand{\bea}{\begin{eqnarray}}
\newcommand{\eea}{\end{eqnarray}}

\catcode`@=12

\begin{document}
\thispagestyle{empty}
\begin{center}
{\Large\bf
{Neutron Star Cooling via Axion Emission by Nucleon-Nucleon Axion 
Bremsstrahlung}}\\
\vspace{1cm}
{{\bf Avik Paul} \footnote{email: avik.paul@saha.ac.in},
{\bf Debasish Majumdar} \footnote{email: debasish.majumdar@saha.ac.in}}\\
\vspace{0.25cm}
{\normalsize \it Astroparticle Physics and Cosmology Division,}\\
{\normalsize \it Saha Institute of Nuclear Physics, HBNI} \\
{\normalsize \it 1/AF Bidhannagar, Kolkata 700064, India}\\
{{\bf Kamakshya Prasad Modak} \footnote{email: kamakshya.modak@gmail.com }}\\
{\normalsize \it Department of Physics, Brahmananda Keshab Chandra College,}\\
{\normalsize \it 111/2, B. T. Road, Kolkata 700108, India}\\
\vspace{1cm}
\end{center}
\begin{abstract}
Neutron stars generally cools off by the emission of gamma rays
and neutrinos. But
axions can also be produced inside a neutron star by the process of 
nucleon-nucleon axion bremsstrahlung. The escape of these axions adds to the
cooling process of neutron star. We explore the nature of cooling 
of neutron stars including the axion emission and compare our result 
with the scenario when the neutron star is cooled by only the 
emission of gamma rays and neutrinos. 
In our calculations we consider 
both the 
degenerate and non-degenerate limits 
for such 
axion energy loss rate and the resulting variation of luminosity with time and variation of surface temperature with time
of the neutron star.
In short the thermal evolution of a
neutron star is studied with three neutron star masses (1.0, 1.4, 1.8 solar masses) and by including the effect of 
axion emission for different axion masses ($m_{a}=10^{-5}\rm{eV}, 10^{-3}\rm{eV}, 10^{-2}\rm{eV}$)
and compared with the same when the axion emission is not considered. We compared theoretical cooling curve with the observational data of three pulsars PSR B0656+14, Geminga and PSR B1055-52 and finally give an upper bound on axion mass limits $m_{a}\leq10^{-3}$ eV which implies that the axion decay constant $f_{a}\geq0.6\times10^{10} \hspace{1mm} \rm{GeV}$.
\end{abstract}
\newpage
\section{Introduction}
A neutron star (NS) \cite{Baym:2005de, Prakash:2014tva} with a typical radius of 10-12 km and generally having a mass range of 1-2 solar mass ($\rm{{M}}_{\odot}$) is formed as an aftermath of a massive supernova explosion. A neutron star can be cooled principally by the emission of photons and neutrinos. It is also conjectured that the emission of axion from a neutron star may contribute to its cooling process in addition to photon and neutrino emissions. Axions \cite{Wilczek:1977pj, Weinberg:1977ma} are pseudo Nambu-Goldstone bosons which are introduced to circumvent the strong CP problem \cite{Peccei:2006as} that refers to the presence of CP violation term in QCD Lagrangian \cite{tHooft:1976rip} arising from the non-Abelian nature of QCD gauge symmetry. The Peccei-Quinn (PQ) solution \cite{Peccei:2006as} of strong CP problem results in the prediction of new particle namely axion. The axion which is a Goldstone boson that arises out of the PQ solution where an anomalous chiral symmetry U(1)$_{\rm A}$ \cite{Peccei:1977hh} is introduced and is spontaneously broken at the PQ energy scale, is an interesting candidate in addressing several aspects of cosmology and particle physics such as dark matter in the Universe. Since the axion have tiny couplings with photons, nucleons and electrons, they can be produced inside a neutron star through the nucleon-nucleon axion bremsstrahlung process $ N + N \rightarrow N + N + a $, where $N$ is a nucleon (proton or neutron) and $``a"$ denotes axion. As mentioned earlier, while the energy loss is considered to be mostly due to photon and neutrino emission, the emission of axion from the late stages of the star such as the supernova or neutron star can also contribute in considerable measure to the process of their cooling. Here we have considered that neutron star is the source of the axions. But axions can be emitted from the other astrophysical sources like the sun, white dwarf, supernova, red giant, globular clusters etc. by the mechanisms such as Primakoff, electron bremsstrahlung, Compton processes etc. Axion mass is also constrained by the experimental \cite{Ringwald:2016}, astrophysical and cosmological limits. These limits indicate that axion have very low mass $m_{a}\lesssim 10 \hspace{1mm}\rm {meV}$ \cite{Hagmann:2010}. Cosmological data constrain the axion mass as $m_{a}< 0.4-1.0 \hspace{1mm}\rm {eV}$ at the $95\%$ statistical CL \cite{Melchiorri:2007}. Solar axions are produced mainly by the Primakoff process. The CAST experiment put an upper limit on the solar axion mass to be $m_{a}\lesssim 0.02 \hspace{1mm}\rm {eV}$ \cite{Zioutas:2005}. In case of globular-clusters, the limits on axion decay constant is given as $f_{a}> 2.3\times 10^{7} \hspace{1mm}\rm {GeV}$ ($m_{a}< 0.3\hspace{1mm}\rm {eV}$) when KSVZ model \cite{Cadamuro:2012rm, Leskinen:2016tfu} is considered while for DFSZ model \cite{Cadamuro:2012rm}-\cite{Berezhiani:1992rk} $f_{a}> 0.8\times 10^{7} \hspace{1mm}\rm {GeV}$ ($m_{a}< 0.7\hspace{1mm}\rm {eV}$) \cite{Hagmann:2010}. But for the case of white dwarf, the constraints on $f_{a}$ is given by $f_{a}> 1.3\times 10^{9}\hspace{1mm} \rm {GeV}\hspace{1mm} {\cos^2{\beta}} $ ($m_{a}< 4.5\hspace{1mm} \rm {meV}\hspace{1mm}/ {\cos^2{\beta}} $) \cite{Hagmann:2010} where $\tan {\beta}$ is the ratio of two Higgs vacuum expectation values. These parameters are useful for explaining the white dwarf cooling by the axion emission \cite{Raffelt:2006cw}. The constraint on axion mass limit for the red giant case is given as (\cite{Raffelt:1995}) $f_{a}\gtrsim 6.6\times 10^{8} \hspace{1mm}\rm {GeV}\hspace{1mm} {\cos^2{\beta}} $ ($m_{a}< 0.009\hspace{1mm} \rm {eV}/\hspace{1mm} {\cos^2{\beta}} $) in the DFSZ model. Raffelt \cite{Raffelt:2006cw} also derived similar constraints from the measured duration of the neutrino signal of the Supernova SN 1987A $f_{a}\gtrsim 4\times 10^{8} \hspace{1mm}\rm {GeV}$ ($m_{a}\lesssim 16\hspace{1mm}\rm {meV}$). Supernova energy loss arguments \cite{Raffelt:2006cw} can be explained by the aforementioned axion parameters. The astrophysical and cosmological axion mass limits are however summarised in Fig. 3 of Ref. \cite{Raffelt:2006cw}.

We consider the bremsstrahlung production of axion in the interior of the neutron star. The energy loss at the later stage of a star can be addressed for two conditions, namely degenerate and non-degenerate limits. While the degenerate condition is applied to only neutron stars, the non-degenerate one is for the accretion disc. In earlier works related to such energy loss through axion, Sedrakian \cite{Sedrakian:2015krq} and Umeda $\it{et}$ $\it{al}$ \cite{Umeda:1997da} considered only the degenerate scenario but in this work we consider both degenerate and non-degenerate cases. We performed a detailed study in this regard and obtain the energy loss of neutron star as a function of time as also the variation of temperature as a function of time for the case when axion emission contributes to the cooling in addition to neutrinos and photons. We then compare our results when axion is not included. We find that for the mass of emitted axions $\sim 10^{-5}\rm{eV}$ and higher, the contribution to the neutron star cooling rate differs considerably from that when no axion emission occurs.  

The paper is organised as follows. In Section 2, we briefly describe 
the formalism that includes the cooling process of neutron star and 
axion production and emission from such a neutron star. The expressions for energy loss rate for both the degenerate and non-degenerate limits are also given in this section. Section 3 elaborates the calculations and results while we furnish some summary and discussions in Section 4.       
\section{Formalism}
In this section we furnish a brief account of axion production and emission from the neutron star. We also furnish the cooling of the star when the effect of axion emission is included.  

\subsection{Axion Emission via Nucleon Nucleon Axion Bremsstrahlung process}
Axions can be emitted from an NS via the nucleon-nucleon axion bremsstrahlung process $ N + N \rightarrow N + N + a $, where $N$ refers to a proton $p$ or a neutron $n$ and $``a"$ denotes axion. In this process nucleons are interacting via one-pion exchange (OPE) potential.
\subsubsection{Matrix Elements for the process $ N + N \rightarrow N + N + a $}
The interaction Hamiltonian for the interaction of axions with the nucleons can be written as \cite{Raffelt:1996wa}
\begin{equation} \label{eq:1}
\mathcal{H}_{int}=-\dfrac{C_{N}}{2f_{a}}\overline{\psi}_{N}\gamma_{\mu}\gamma_{5}\psi_{N}\partial^{\mu}a
\end{equation}
where $f_{a}$ is the PQ energy scale for axions, $C_{N}$ is a dimensionless model dependent coupling constant of order unity, the $\psi_{N}$ represents the nucleon Dirac fields and $``a"$ is the axion field.
The possible Feynman diagrams related to this process is given in Ref. \cite{Brinkmann:1988vi}. There are two kinds of diagrams. One (four of them) refers to the processes where the axion is attached to each nucleon line while the other four diagrams are for the exchange processes ($ N_{3} \leftrightarrow N_{4}$) where the axion is attached to each nucleon line and $N_{1}, N_{2}, N_{3}$ and $N_{4}$ denote the four nucleons which take part in bremsstrahlung process. For the processes, $nn \rightarrow nn + a$ and $ pp \rightarrow pp + a$ (``pure processes") the spin-summed squared matrix element is of the form \cite{Brinkmann:1988vi, Raffelt:1993ix}
\begin{equation} \label{eq:2}
\sum_{spins}{{\mid \mathcal{M} \mid}^2_{NN}}=\dfrac{16({4\pi})^{3}\alpha_{\pi}^{2}\alpha_{a}}{3{m_{N}}^{2}}\bigg[\bigg(\dfrac{\textbf{K}^{2}}{\textbf{K}^{2}+m_{\pi}^{2}}\bigg)^{2}+\bigg(\dfrac{\textbf{l}^{2}}{\textbf{l}^{2}+m_{\pi}^{2}}\bigg)^{2}+\dfrac{\textbf{K}^{2}\textbf{l}^{2}-3\bigg(\textbf{K}.\textbf{l}\bigg)^{2}}{(\textbf{K}^{2}+m_{\pi}^{2})(\textbf{l}^{2}+m_{\pi}^{2})}\bigg]
\end{equation}
where $m_{\pi}$ is the pion mass and $m_{N}$ is the nucleon mass. The axion-nucleon ``fine-structure constant" $\alpha_{a}\equiv(C_{N}m_{N}/f_{a})^{2}/4\pi=g_{aN}^{2}/4\pi$, where $g_{aN}=(C_{N}m_{N}/f_{a})$ is the axion-nucleon coupling constant. The quantity pion-nucleon ``fine structure constant" is given by $\alpha_{\pi}\equiv(f2m_{N}/m_{\pi})^{2}/4\pi\approx17$ where the pion-nucleon coupling $f\approx1.05$ and $\textbf{K}=\textbf{p}_{2}-\textbf{p}_{4}$ and  $\textbf{l}=\textbf{p}_{2}-\textbf{p}_{3}$ 
with $p_{i}$ ($i=1-4$) are the momenta of the nucleons $N_{i}$. 
For the ``mixed" process $np \rightarrow np+a$, spin-summed squared matrix element can be written as \cite{Raffelt:1993ix}
\begin{equation} \label{eq:3}
\begin{aligned}
\sum_{spins}{{\mid \mathcal{M} \mid}^2_{np}}=\dfrac{256{\pi}^{2}\alpha_{\pi}^{2}}{3{m_{N}}^{2}}\dfrac{(g_{an}+g_{ap})^{2}}{4}\bigg[2\bigg(\dfrac{\textbf{l}^{2}}{\textbf{l}^{2}+m_{\pi}^{2}}\bigg)^{2}-\dfrac{4\bigg(\textbf{K}.\textbf{l}\bigg)^{2}}{(\textbf{K}^{2}+m_{\pi}^{2})(\textbf{l}^{2}+m_{\pi}^{2})}\bigg]\\
+\dfrac{256{\pi}^{2}\alpha_{\pi}^{2}}{3{m_{N}}^{2}}\dfrac{(g_{an}^{2}+g_{ap}^{2})}{2}\bigg[\bigg(\dfrac{\textbf{K}^{2}}{\textbf{K}^{2}+m_{\pi}^{2}}\bigg)^{2}+2\bigg(\dfrac{\textbf{l}^{2}}{\textbf{l}^{2}+m_{\pi}^{2}}\bigg)^{2}\\
+2\dfrac{\textbf{K}^{2}\textbf{l}^{2}-\bigg(\textbf{K}.\textbf{l}\bigg)^{2}}{(\textbf{K}^{2}+m_{\pi}^{2})(\textbf{l}^{2}+m_{\pi}^{2})}\bigg].
\end{aligned}
\end{equation}
\subsubsection{Energy Loss Rate Expression}
The axion energy-loss rate per unit volume is given by \cite{Raffelt:1996wa}
\begin{equation} \label{eq:4}
\begin{aligned}
Q_{a}=\int{\dfrac{d^{3}\textbf{K}_{a}}{2\omega_{a}(2\pi)^{3}}}\omega_{a}\int{\prod_{i=1}^{4}\dfrac{d^{3}\textbf{P}_{i}}{2E_{i}(2\pi)^{3}}}f_{1}f_{2}(1-f_{3})(1-f_{4})\\
\times(2\pi)^{4}\delta^{4}(P_{1}+P_{2}-P_{3}-P_{4}-K_{a})S\sum_{spins}{{\mid \mathcal{M} \mid}^2_{NN}}
\end{aligned}
\end{equation}
where $P_{1}$ and $P_{2}$ are the four-momenta of the initial-state nucleons, $P_{3}$ and $P_{4}$ are the four-momenta of the final states nucleons and $K_{a}$ is the four-momentum of the axion. A factor $S$ is introduced to accounts for the identical particles in the initial and final states and it takes the value $S = 1/4$ for pure processes and $S = 1$ for mixed processes. In Eq. \ref{eq:4} $f_{i}$'s are the occupation numbers for the nucleons $N_{i}$'s. In this case, it is assumed that the axions escape freely so that a Bose stimulation factor $(1+f_{a})$ as well as axion absorption factors are neglected.

In the following, we furnish the simplified expressions for energy loss rate $Q_{a}$ for both degenerate and non-degenerate limits.
\subsubsection{Non-degenerate Limit}
For simplicity, we first neglect the pion mass contribution in Eqs. \ref{eq:2} and \ref{eq:3}. With this approximation, the squared matrix element reduces to \cite{Greenberg:2000he}
\begin{equation} \label{eq:6}
\sum_{spins}{{\mid \mathcal{M} \mid}^2_{NN}}=\dfrac{256{\pi}^{2}\alpha_{\pi}^{2}}{3{m_{N}}^{2}}\tilde{g}_{NN}^{2}
\end{equation}
with $\tilde{g}_{NN}^{2} \equiv g_{an}^{2}(3-\beta)$ for $nn\rightarrow nn+a$ process, $\tilde{g}_{NN}^{2} \equiv g_{ap}^{2}(3-\beta)$ for $pp\rightarrow pp+a$ process, and \\$\tilde{g}_{NN}^{2} \equiv \left(\dfrac{g_{an}+g_{ap}}{2}\right)^{2}(2-4\beta/3)+\dfrac{g_{an}^{2}+g_{ap}^{2}}{2}(5-2\beta/3)$ for $np\rightarrow np+a$ process. The effective coupling $\tilde{g}_{NN}$ is given by $\tilde{C}_{N}m_{N}/f_{a}$ and $\beta \equiv 3\langle(\hat{\textbf{K}}.\hat{\textbf{l}})^{2}\rangle.$ In the non-degenerate limit the numerical value of $\beta$ is 1.3078 \cite{Raffelt:1993ix}. The axion-nucleon coupling $\mid\tilde{C}_{N}\mid$ can be found numerically as $\mid\tilde{C}_{N}\mid =0.013$ for $nn\rightarrow nn+a$ process,
$\mid\tilde{C}_{N}\mid =0.442$ for $pp\rightarrow pp+a$ process and
$\mid\tilde{C}_{N}\mid =0.495$ for $np\rightarrow np+a$ process \cite{Greenberg:2000he}.  Introducing a ``fudge factor" $\xi(T)$ to account for the pion mass effects and using Eq. \ref{eq:6} in Eq. \ref{eq:4} yields the total energy loss rate per unit volume as \cite{Greenberg:2000he}
\begin{equation} \label{eq:7}
\begin{aligned}
Q_{a}^{ND}=\dfrac{\xi(T)}{280}\dfrac{n_{B}^{2}T^{7/2}}{m_{N}^{5/2}\pi^{7/2}}\Big(Y_{n}^{2}\sum_{spins}{\mid \mathcal{M} \mid}^2_{nn}+Y_{p}^{2}\sum_{spins}{\mid \mathcal{M} \mid}^2_{pp}+Y_{n}Y_{p}\sum_{spins}{\mid \mathcal{M} \mid}^2_{np}\Big)\\
=\dfrac{32}{105}\xi(T)\dfrac{\alpha_{\pi}^{2}n_{B}^{2}T^{7/2}}{m_{N}^{9/2}\pi^{3/2}}\Big(Y_{n}^{2}\tilde{g}_{nn}^{2}+Y_{p}^{2}\tilde{g}_{pp}^{2}+4Y_{n}Y_{p}\tilde{g}_{np}^{2}\Big)\hspace{3.8cm}\\
=\dfrac{32}{105}\xi(T)\dfrac{\alpha_{\pi}^{2}n_{B}^{2}T^{7/2}}{m_{N}^{9/2}\pi^{3/2}}g_{ND}^{2}\hspace{8.3cm}
\end{aligned}
\end{equation}
where $g_{ND}$ is the total effective axion-nucleon coupling constant for the non-degenerate limit, $Y_{p}$ is the proton number fraction and $Y_{n}$ is the neutron number fraction. With $Y_{p} \approx 0.1$, $Y_{n} \approx 0.9$, $\xi(T)\approx 0.5$, and using the above $\mid\tilde{C}_{N}\mid $ values, $\tilde{g}_{NN}^{2}$ expressions and the relation $m_{a}=6\rm{\mu eV} \left(\frac{10^{12}\rm{GeV}}{f_{a}}\right)$ \cite{Berenji:2016jji}, the values of $g_{ND}$ and $C_{N}^{ND}$ can be found numerically as
\begin{equation} \label{eq:8}
g_{ND}=4.71\times 10^{-8}\hspace{1mm}\Big(\dfrac{m_{a}}{eV}\Big)\hspace{1cm} \rm{and} \hspace{1cm}C_{N}^{ND}=0.30.
\end{equation}
With the above expression for $g_{ND}$ the axion energy loss rate per unit volume (Eq. \ref{eq:7}) reduces to
\begin{equation} \label{eq:9}
Q_{a}^{ND}=2.90166\times 10^{31} \hspace{1mm}erg\hspace{1mm}cm^{-3}\hspace{1mm}yr^{-1}\hspace{1mm}T^{3.5}_{9}\hspace{1mm}\rho_{12}^{2}\hspace{1mm}m^{2}_{eV}
\end{equation}
where $m_{eV} \equiv m_{a}/eV $, $T_{9}\equiv T/10^{9} K$, and $\rho_{12} \equiv \dfrac{\rho}{10^{12} \rm{g/{cm}^{3}}}=n_{B}m_{N}$ where $n_{B}$ is the nucleon (baryon) density.
\subsubsection{Degenerate Limit}
In the degenerate limit, Eq. (\ref{eq:6}) simplies to \cite{Greenberg:2000he}
\begin{equation} \label{eq:11}
\sum_{spins}{{\mid \mathcal{M} \mid}^2_{NN}}=\dfrac{256{\pi}^{2}\alpha_{\pi}^{2}}{{m_{N}}^{2}}\tilde{g}_{NN}^{2}.
\end{equation}
Unlike the non-degenerate case, here the parameter $\beta$ is zero.
With $\tilde{g}_{NN}^{2}\equiv g_{an}^{2}$ for $nn\rightarrow nn+a$ process,
$\tilde{g}_{NN}^{2} \equiv g_{ap}^{2}$ for $pp\rightarrow pp+a$ process and
$\tilde{g}_{NN}^{2}\equiv \left({g_{an}^{2}+g_{ap}^{2}}+(g_{an}g_{ap})/3\right)$ for $np\rightarrow np+a$ process.
The numerical values of effective coupling are
$\mid\tilde{C}_{N}\mid=0.01$ for $nn\rightarrow nn+a$ process,
$\mid\tilde{C}_{N}\mid =0.34$ for $pp\rightarrow pp+a$ process and
$\mid\tilde{C}_{N}\mid=0.338$ for $np\rightarrow np+a$ process \cite{Greenberg:2000he}.
In this case the integration in Eq. \ref{eq:4} can be simplified analytically without neglecting the pion masses. Here the pionic contribution $F(u)$ is given by \cite{Raffelt:1996wa}
\begin{equation} \label{eq:12}
\begin{aligned}
F(u)=1-\frac{5u}{6}\hspace{1mm}\arctan(\frac{2}{u})+\frac{u^{2}}{3(u^{2}+4)}+\frac{u^{2}}{6\sqrt{2u^{2}+4}}\\
\times\arctan \Big(\frac{2\sqrt{2u^{2}+4}}{u^{2}}\Big)
\end{aligned}
\end{equation}
where $u=m_{\pi}/p_{F,N}$. {With $u\approx0.32Y_{N}^{-1/3}$ and consider $\rho_{B}\approx 2\rho_{nuc}$, {$F(u)$ can be replaced by $F(Y_{N})$ and the total emission rate is given as \cite{Raffelt:1996wa, Greenberg:2000he}
\begin{equation} \label{eq:13}
\begin{aligned}
Q_{a}^{D}=\dfrac{31}{967680}\Big(\dfrac{3n_{B}}{\pi}\Big)^{1/3}T^{6}\Big(Y_{n}^{1/3}F(Y_{n})\sum_{spins}{\mid \mathcal{M} \mid}^2_{nn}+Y_{p}^{1/3}F(Y_{p})\sum_{spins}{\mid \mathcal{M} \mid}^2_{pp}\\+4Y_{np}^{1/3}F(Y_{np})\sum_{spins}{\mid \mathcal{M} \mid}^2_{np}\Big)\\
=\dfrac{31\pi^{5/3}(3n_{B})^{1/3}\alpha_{\pi}^{2}T^{6}}{3780m_{N}^{2}}\Big(Y_{n}^{1/3}F(Y_{n})\tilde{g}_{an}^{2}+Y_{p}^{1/3}F(Y_{p})\tilde{g}_{ap}^{2}+Y_{np}^{1/3}F(Y_{np})\tilde{g}_{np}^{2}\Big)\hspace{-0.13cm}\\
=\dfrac{31\pi^{5/3}(3n_{B})^{1/3}\alpha_{\pi}^{2}T^{6}}{3780m_{N}^{2}}g_{D}^{2}\hspace{8.2cm},
\end{aligned}
\end{equation}
where $g_{D}$ is the total effective axion-nucleon coupling constant for the degenerate limit. The effective nucleon fraction $Y_{np}$ for the mixed processes are given by \cite{Brinkmann:1988vi}
\begin{equation} \label{eq:14}
Y_{np}^{1/3}=\dfrac{1}{2\sqrt{2}}(Y_{n}^{2/3}+Y_{p}^{2/3})^{1/2}\Big[2-\dfrac{\mid Y_{n}^{2/3}-Y_{p}^{2/3} \mid}{Y_{n}^{2/3}+Y_{p}^{2/3}}\Big].
\end{equation}
For the degenerate case, using the values of nucleon number fractions $Y_{p}=0.01$, $Y_{n}=0.99$ and $Y_{np}=0.06$, we can calculate $F(Y_{n})\approx 0.64, F(Y_{p})\approx 0.12 \hspace{2mm}$ and $F(Y_{np})\approx 0.31$ for the pure and mixed processes. Also using the above values one can attain the effective coupling constants $g_{D}$ and $C_{N}^{D}$ for the degenerate case as
\begin{equation} \label{eq:15}
g_{D}=2.04\times 10^{-8}\hspace{1mm}\Big(\dfrac{m_{a}}{eV}\Big)\hspace{1cm} \rm{and} \hspace{1cm}C_{N}^{D}=0.13.
\end{equation}
Using Eqs. \ref{eq:13} and \ref{eq:15} the axion emission rate per unit volume is obtained as
\begin{equation} \label{eq:16}
Q_{a}^{D}=4.84244\times 10^{30} \hspace{1mm}erg\hspace{1mm}cm^{-3}\hspace{1mm}yr^{-1}\hspace{1mm}T^{6}_{9}\hspace{1mm}m^{2}_{eV}\hspace{1mm}\Big(\dfrac{\rho_{NS}}{\rho_{nuc}}\Big)^{1/3}
\end{equation} 
where $m_{eV} \equiv m_{a}/eV $, $T_{9}\equiv T/10^{9} K$, $\rho_{NS}$ is the density of the neutron star and $\rho_{nuc}$ is the nuclear density.
\subsection{Neutron star cooling}
In the last few years it has been observed that the surface temperature of the neutron star decreases with time. This is the only direct indication of cooling of a neutron star \cite{Pizzochero:2016ffp}.

In Newtonian formulation the energy balance equation for the neutron star is given by \cite{Sedrakian:2015krq}
\begin{equation} \label{eq:18}
\frac{dE_{\rm th}}{dt}=C_{v}\frac{dT}{dt}=
-L_{\nu}(T)-L_{a}(T)-L_{\gamma}(T_{e})+H(T),
\end{equation}
where $E_{\rm th}$ is the thermal energy content of the star, 
$T$ is its internal temperature and $T_{e}$ is the effective
temperature. The quantities $L_{\nu}$ and $L_{a}$ are the neutrino and axion 
luminosities respectively from the bulk of the star
and $L_{\gamma}$ is the luminosity of photons radiated from the
star surface. In Eq. \ref{eq:18} $C_{v}$ is the specific heat of the
core, and the source term $H$ contains all possible 
``heating mechanisms" which, for example, convert magnetic or
rotational energy into heat energy. This could be significant in the 
late time evolution of neutron stars. It is assumed here that
$H(T)=0$. The photon luminosity $L_{\gamma}$ is given by the Stefan-Boltzmann law \cite{Nscool:2010dpl}
\begin{equation} \label{eq:19}
L_{\gamma}=S\hspace{1mm}T^{2+4\alpha}=4\pi\hspace{1mm}\sigma\hspace{1mm}R^{2}\hspace{1mm}T_{e}^{4}
\hspace{2mm}.
\end{equation}
The above relation is obtained using $\hspace{2mm} T_{e}\propto T^{0.5+\alpha} \hspace{1mm}(\alpha\ll 1)$,
where $\sigma$ is the Stefan-Boltzmann constant, and $R$ is the radius of the star. In the present work we use the NSCool code \cite{Nscool:2010dp} for calculating the axion luminosity. Various neutrino processes are also 
involved in the cooling of neutron stars \cite{Page:2005fq}. Dominant neutrino emitting processes are direct Urca processes and modified
Urca processes. The direct Urca processes are $n\rightarrow p+e^{-
}+\overline{\nu}_{e}$ (beta decay) and $p+e^{-}\rightarrow n+\nu_{e}$ (electron capture), which are only possible in neutron stars if the proton fraction exceeds a critical threshold. The above two neutrino emitting
processes are fast and the dependence of luminosity on temperature is given by the relation $L_{\nu}^{fast}\propto
T^{6}_{9}$. In case the proton fraction is below the threshold, the dominant neutrino emission process, a variant of the direct Urca process, namely the  modified Urca process, is a second-order process. The modified processes
are
\bea \label{eq:20}
n+n\rightarrow n+p+e^{-}+\overline{\nu}_{e}, && 
n+p+e^{-}\rightarrow n+n+\nu_{e}\,\, \nonumber \\ \
{\rm neutron}\,\, {\rm branch} \nonumber \\
p+n\rightarrow p+p+e^{-}+\overline{\nu}_{e}, &&
p+p+e^{-}\rightarrow p+n+\nu_{e}\,\, \ \nonumber \\
{\rm proton}\,\, {\rm branch.} 
\eea
These neutrino emitting processes are slow and the dependence of luminosity on temperature is given by the relation
$L_{\nu}^{slow}\propto T^{8}_{9}$.
The other neutrino emitting processes are electron-positron pair annihilation, plasmon decay, electron synchrotron, photoneutrino emission, electron-nucleus bremsstrahlung, cooper pairing of neutrons, neutron-neutron bremsstrahlung, neutron-nucleus bremsstrahlung \cite{Nscool:2010dpl}.

\section{Calculations and Results}
As mentioned earlier, we use NSCool numerical code \cite{Nscool:2010dp} and include axion  energy loss rate for both degenerate and non-degenerate cases. We adopt Akmal Pandharipande Ravenhall (APR) equation of state (EoS) for our work. This equation of state deals with nuclear degrees of freedom only by which the fast cooling processes are avoided. Moreover, APR does not include non-nucleonic degrees of freedom. In this EoS the two-nucleon interaction, namely Argonne $v_{18}$ \cite{Akmal:1998cf} has been incorporated and the boost corrections to the two-nucleon interaction are also considered. This ensures the relativistic effect in the EoS. In addition, three-nucleon interactions are also considered in the nuclear Hamiltonian. This enables an increase in mass limit of the neutron star to $\sim$ 2.2 ${M}_{\odot}$. For demonstrating the effect of axion emission in the cooling process of neutron star, we consider neutron stars of three different masses namely 1.0${M}_{\odot}$ (light), 1.4${M}_{\odot}$ (intermediate) and 1.8${M}_{\odot}$ (massive) and obtain the variation of their luminosities with surface temperature and time for the cases including axion emission in the cooling process and without axion emission. We also consider three axion masses ($m_{a}$) namely $m_{a}=10^{-5}\rm{eV}$, $10^{-3}\rm{eV}$ and $10^{-2}\rm{eV}$, in the calculation. The energy loss rate due to axion emission increases with the increase of axion mass because of $m_{a}^{2}$ factor in the energy loss rate formula (Eqs. \ref{eq:9}, \ref{eq:16}) which implies fast cooling of neutron star and fast cooling occurs for the increase of neutron star masses. In Figs. \ref{fig:3}$-$\ref{fig:5} we show how the neutron star luminosities vary with time due to neutron star cooling when axion cooling is included along with the cooling due to neutrino and gamma emission and compared with the case when axion emission is not considered for neutron star masses 1.0${M}_{\odot}$, 1.4${M}_{\odot}$ and 1.8${M}_{\odot}$ respectively. The left panel of Figs. \ref{fig:3}$-$\ref{fig:5} are for degenerate case while the right panels are for non-degenerate case. In Figs. \ref{fig:3}$-$\ref{fig:8} we include observational results for three pulsars, namely PSR B0656+14, Geminga and PSR B1055-52. The results are shown by points in all the figures.

It can be observed from Figs. \ref{fig:3}$-$\ref{fig:5} that heavier the axion masses further away are the luminosity-time plots from the observational data. For the degenerate category, the variation of luminosities with time almost coincide with the results when no axion mass is considered. This is true for all the three chosen masses of neutron stars. It can also be observed from the left panel of Fig. \ref{fig:5} that for the 1.8${M}_{\odot}$ neutron star, the observational data for the luminosities for the pulsars PSR B0656+14 and Geminga fairly agree with the case when axion mass is $10^{-5}\rm{eV}$. In addition, the luminosity with axion mass $10^{-3}\rm{eV}$ is within the error bar of the Geminga pulsar observational data. For the non-degenerate case however, the cooling due to axion emission appears to cause the depletion of both luminosities and temperatures more rapidly in comparison to degenerate cases, at the region of more advanced stages (higher time) of the neutron star. This trend becomes more prominent for lower neutron star mass. For example, neutron star mass of 1.8${M}_{\odot}$ the cooling curve (Fig. \ref{fig:5}, right panel) with axion emission for axion mass $10^{-5}\rm{eV}$ barely differs from that without axion emission after $t\sim 10^{5}$ years but this difference increases when neutron star mass is 1.4${M}_{\odot}$ or 1.0${M}_{\odot}$ (Fig. \ref{fig:3}, right panel and Fig. \ref{fig:4}, right panel). For the other two chosen axion masses ($10^{-3}\rm{eV}$,  $10^{-2}\rm{eV}$), the cooling curve however, shows much rapid decrease of luminosity. Similar trends are also observed for the temperature vs. time plots for non-degenerate cases but the temperature in this case falls off more rapidly for all the chosen axion masses (Figs. \ref{fig:6}$-$  \ref{fig:8}).  

Calculations are also done to obtain the variations of neutron star temperature with time for the same set of three neutron star masses and axion masses as in the previous case (Figs. \ref{fig:3}$-$\ref{fig:5}). These are ploted in Figs. \ref{fig:6}$-$\ref{fig:8}. Here too the left panel of Figs. \ref{fig:6}$-$\ref{fig:8} are for the degenerate limit while the right panels of these Figures show the results for non-degenerate limit. Once again, one notices that for the degenerate case the results for axion mass $m_{a}=10^{-5}\rm{eV}$ almost coincide with that for the case when no axion cooling is considered. It is also to be noted that the neutron star temperature calculations are agree with the observational data for pulsars PSR B0656+14 and Geminga at least upto their error bars for all the three neutron star masses considered here. Similar trends are also obtained for non-degenerate case. These results and their comparison with the observational data of three pulsars can be indicative of the fact that the mass of the axion would be below $10^{-3}\rm{eV}$.

We also make a direct comparison between the degenerate case and the non-degenerate case in Figs. \ref{fig:9}$-$\ref{fig:12}. In Fig. \ref{fig:9}, the results for degenerate and non-degenerate limits are shown by calculating and ploting the variation of luminosities with time for both the limits when the axion mass is $10^{-5}\rm{eV}$. The results are shown for all the three neutron star masses, namely 1.0${M}_{\odot}$, 1.4${M}_{\odot}$ and 1.8${M}_{\odot}$ considered in this work. In Fig. \ref{fig:10} we show similar results but for axion mass $m_{a}=10^{-3}\rm{eV}$. In both Figs. \ref{fig:9} and \ref{fig:10} it can be seen that while at early times results for both the degenerate and non-degenerate limits appear to almost overlap, at later times this difference decreases with the increase in neutron star mass. Therefore for very low axion mass ($\sim10^{-5}\rm{eV}$) the degenerate and non-degenerate cases appears to be increasingly indistinguishable as the neutron star becomes more and more massive. In contrast, we observe from Fig. \ref{fig:10} that with the increase in axion mass (in this case $\sim10^{-3}\rm{eV}$) the difference in results for degenerate and non-degenerate cases, although decreases with increase of neutron star mass, does not tend to vanish at high neutron star mass. Similar plots as in Figs. \ref{fig:9} and \ref{fig:10} but for the variation of temperature and time are shown in Figs. \ref{fig:11} and \ref{fig:12}. While Fig. \ref{fig:11} is for axion mass $10^{-5}\rm{eV}$, in Fig. \ref{fig:12} we plot the result temperature vs. time when the axion mass is $10^{-3}\rm{eV}$. The results are shown all the three neutron star masses considered.
Similar trends regarding the variations of differences between the degenerate and non-degenerate cases are also noticed when temperature is varied with time.

\begin{figure}
\includegraphics[width=8.5cm,height=8.5cm]{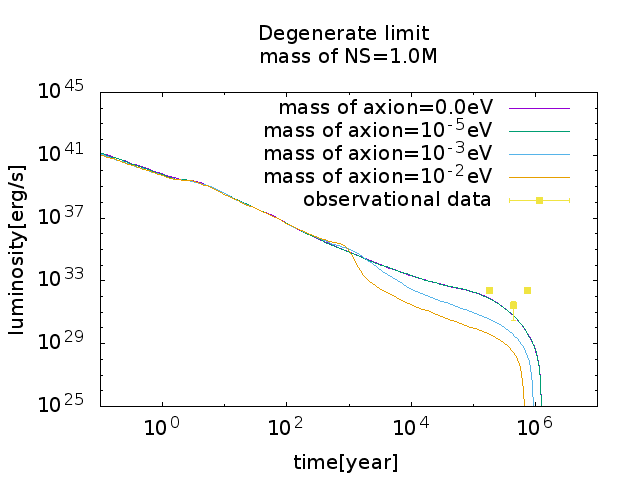}
\includegraphics[width=8.5cm,height=8.5cm]{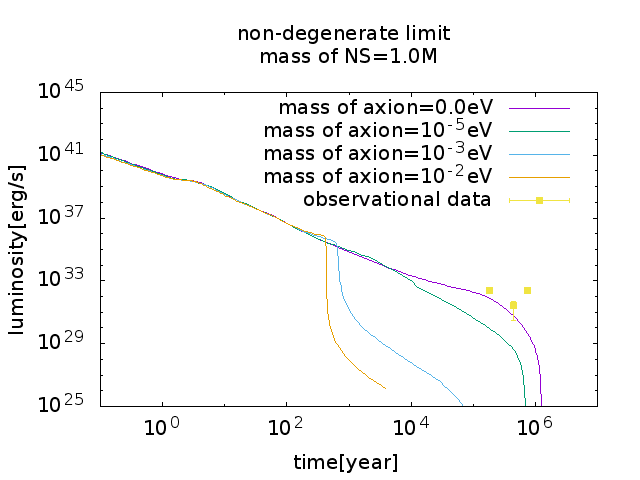}
\caption{Luminosity ($L$) vs. time ($t$) graph for both degenerate (left panel) and non-degenerate (right panel) limits with $ \rm{M}=1.0\rm{M}_{\odot}$ and $m_{a}=0\rm{eV}, 10^{-5}\rm{eV}, 10^{-3}\rm{eV}, 10^{-2}\rm{eV}$ (from top to bottom). The observational data for three pulsars PSR B0656+14, Geminga and PSR B1055-52 are shown by dots with error bars from left to right.}
\label{fig:3}
\end{figure}
\begin{figure}
\includegraphics[width=8.5cm,height=8.5cm]{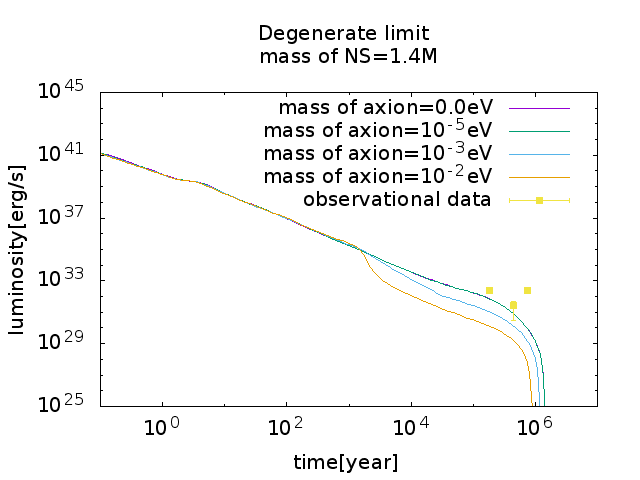}
\includegraphics[width=8.5cm,height=8.5cm]{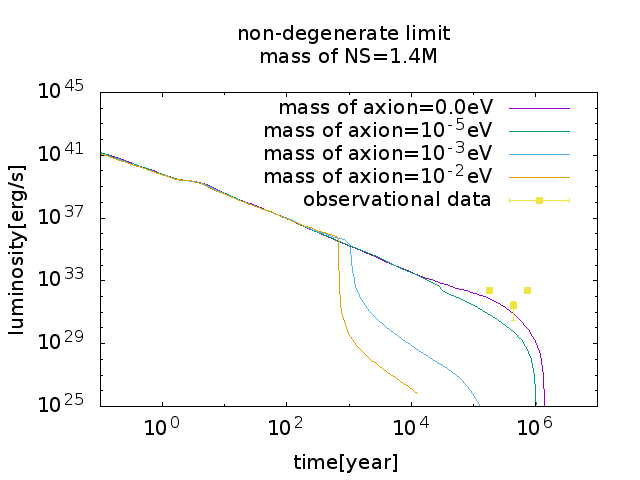}
\caption{Same as in Fig. \ref{fig:3} but for $ \rm{M}=1.4\rm{M}_{\odot}$}
\label{fig:4}
\end{figure}
\begin{figure} 
\includegraphics[width=8.5cm,height=8.5cm]{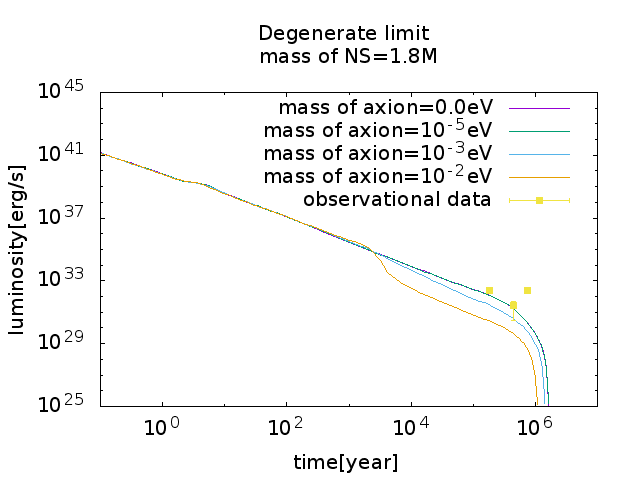}
\includegraphics[width=8.5cm,height=8.5cm]{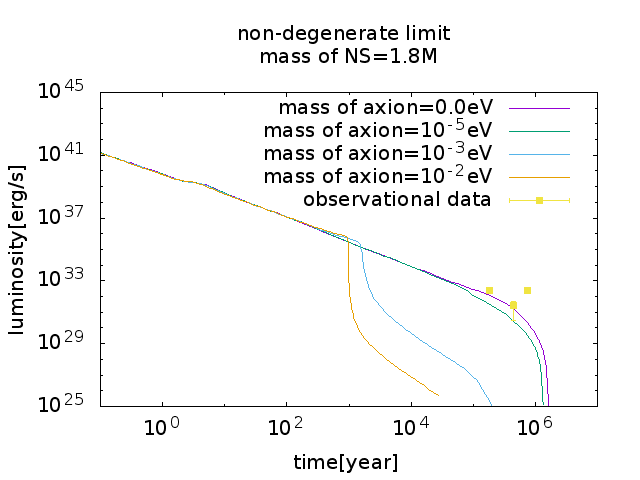}
\caption{Same as in Fig. \ref{fig:3} but for $ \rm{M}=1.8\rm{M}_{\odot}$}
\label{fig:5}
\end{figure}
\begin{figure} 
\includegraphics[width=8.5cm,height=8.5cm]{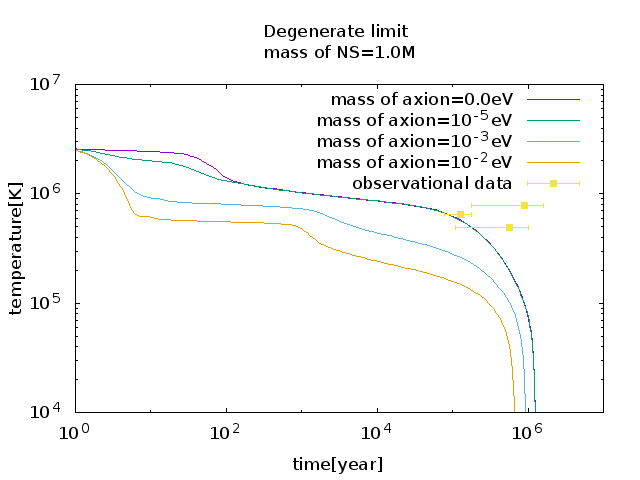}
\includegraphics[width=8.5cm,height=8.5cm]{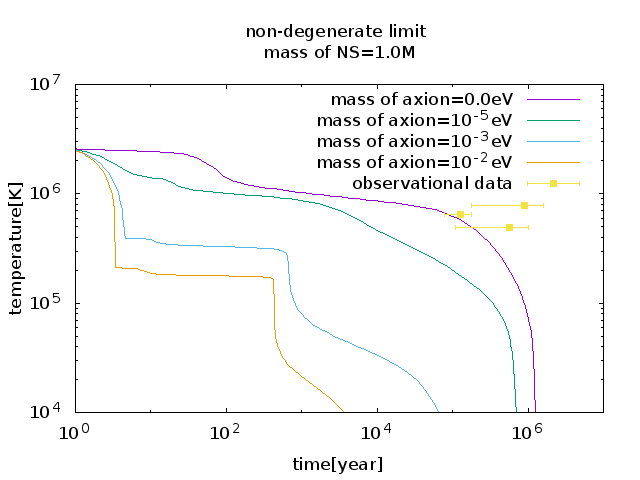}
\caption{Surface temperature ($T$) vs. time ($t$) graph for both degenerate (left panel) and non-degenerate (right panel) limits with $ \rm{M}=1.0\rm{M}_{\odot}$ and $m_{a}=0\rm{eV}, 10^{-5}\rm{eV}, 10^{-3}\rm{eV}, 10^{-2}\rm{eV}$ (from top to bottom).  The observational data for three pulsars PSR B0656+14, Geminga and PSR B1055-52 are shown by dots with error bars from left to right.}
\label{fig:6}
\end{figure}
\begin{figure} 
\includegraphics[width=8.5cm,height=9cm]{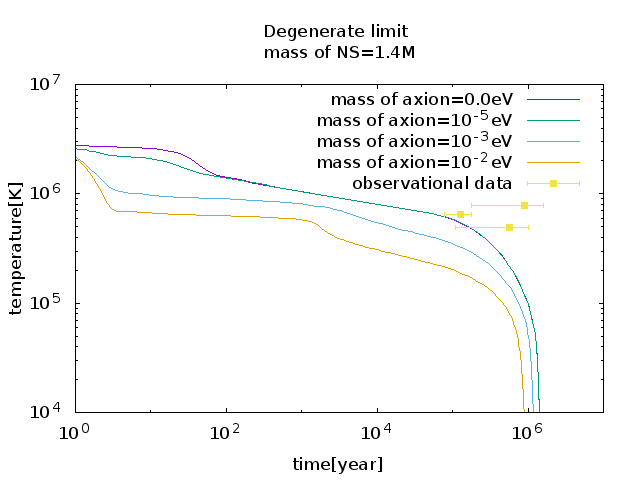}
\includegraphics[width=8.5cm,height=9cm]{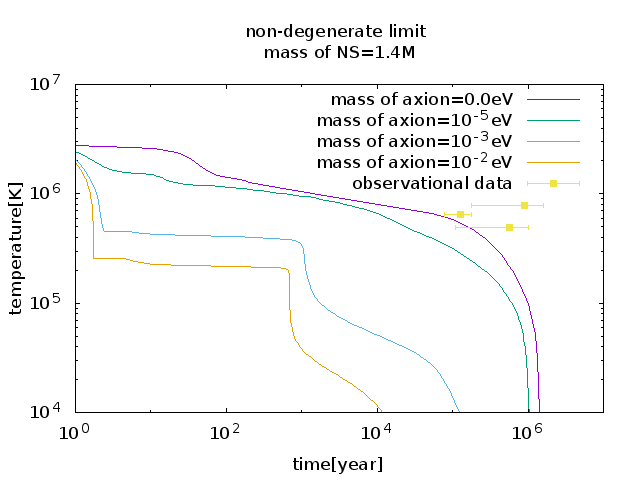}
\caption{Same as in Fig. \ref{fig:6} but for $ \rm{M}=1.4\rm{M}_{\odot}$}
\label{fig:7}
\end{figure}
\begin{figure} 
\includegraphics[width=8.5cm,height=9.5cm]{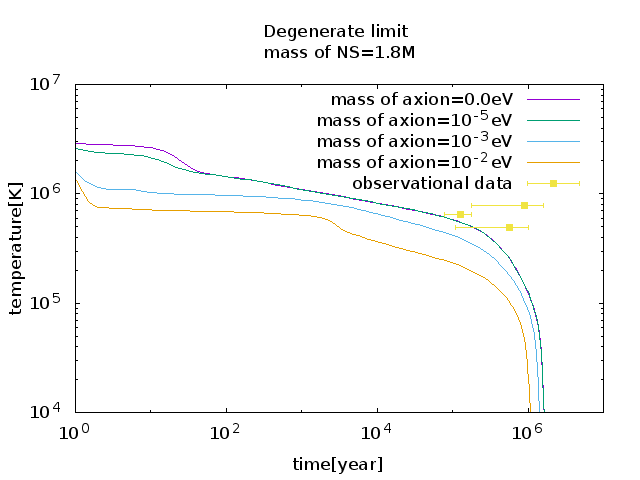}
\includegraphics[width=8.5cm,height=9.5cm]{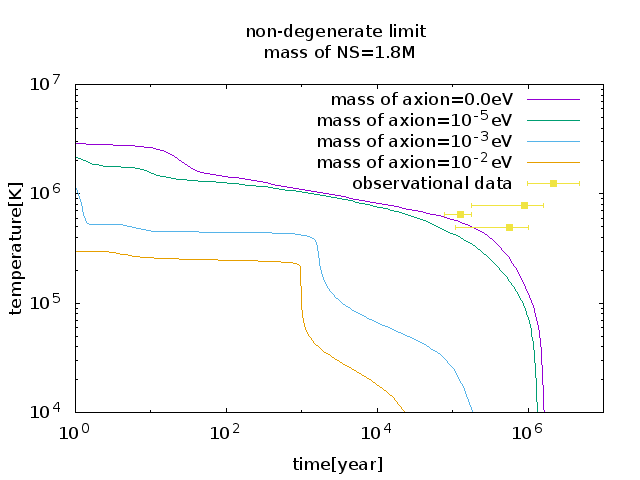}
\caption{Same as in Fig. \ref{fig:6} but for $ \rm{M}=1.8\rm{M}_{\odot}$}
\label{fig:8}
\end{figure}
\begin{figure} 
\includegraphics[width=5.5cm,height=7cm]{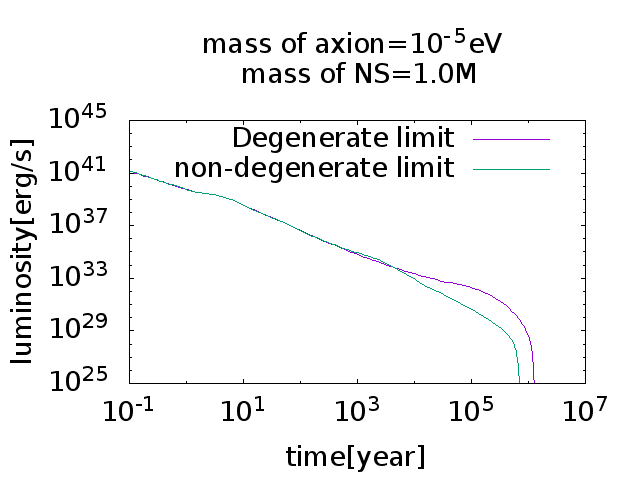}
\includegraphics[width=5.5cm,height=7cm]{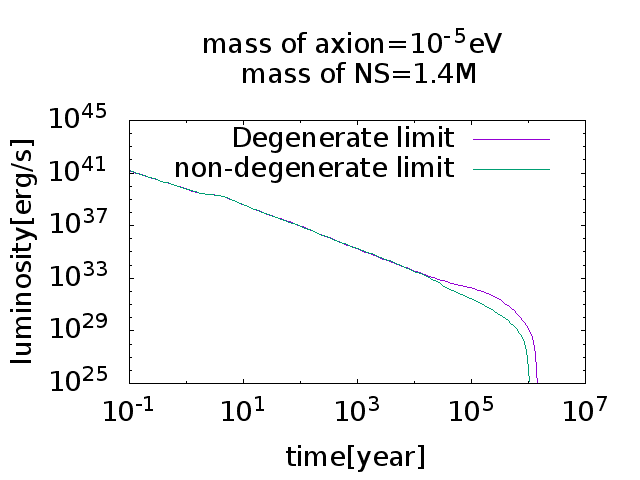}\includegraphics[width=5.5cm,height=7cm]{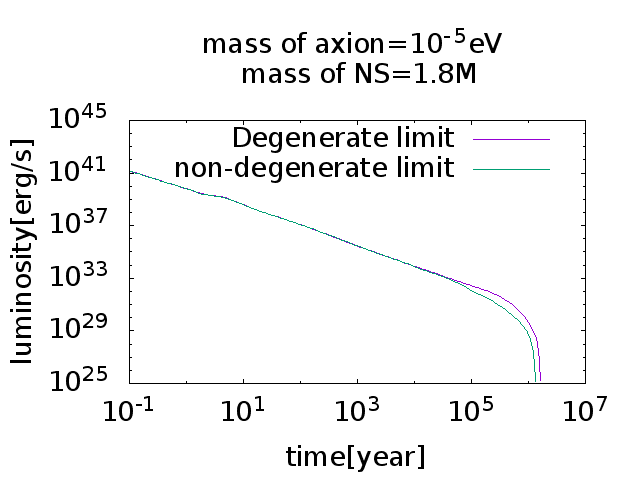}
\caption{Comparison between degenerate and non-degenerate limits for $L$ vs. $t$ plot with $m_{a}$=$10^{-5}\rm{eV}$ and $ \rm{M}=1.0\rm{M}_{\odot}$, $ \rm{M}=1.4\rm{M}_{\odot}$ and $ \rm{M}=1.8\rm{M}_{\odot}$ (from left to right).}
\label{fig:9}
\end{figure}
\begin{figure} 
\includegraphics[width=5.5cm,height=7cm]{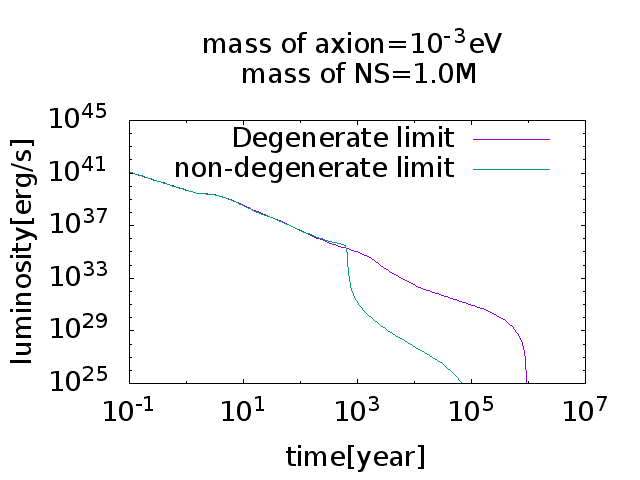}
\includegraphics[width=5.5cm,height=7cm]{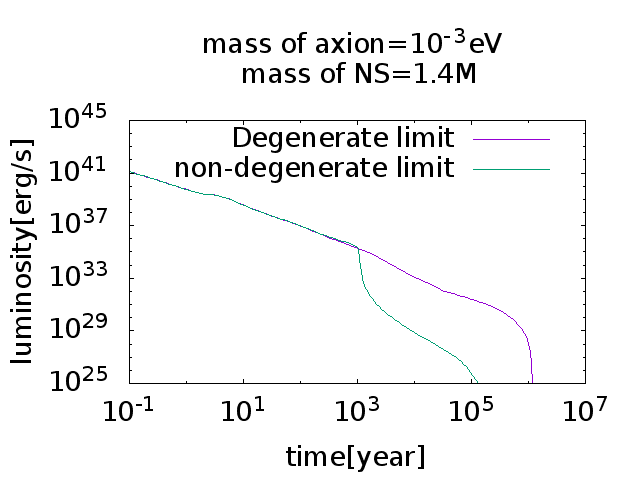}\includegraphics[width=5.5cm,height=7cm]{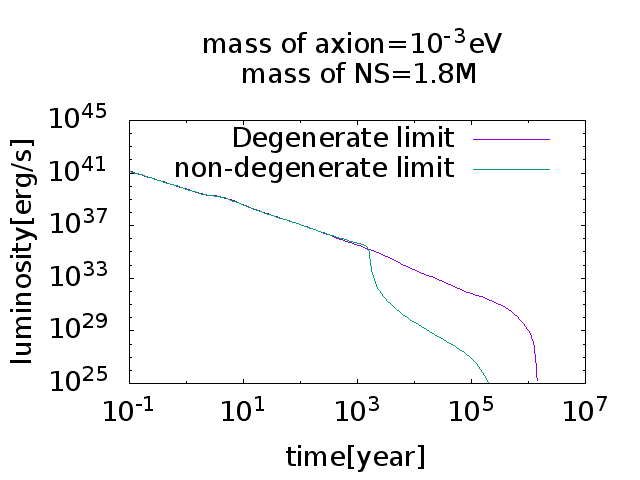}
\caption{Same as in Fig. \ref{fig:9} but for $m_{a}$=$10^{-3}\rm{eV}$.}
\label{fig:10}
\end{figure}
\begin{figure} 
\includegraphics[width=5.5cm,height=7cm]{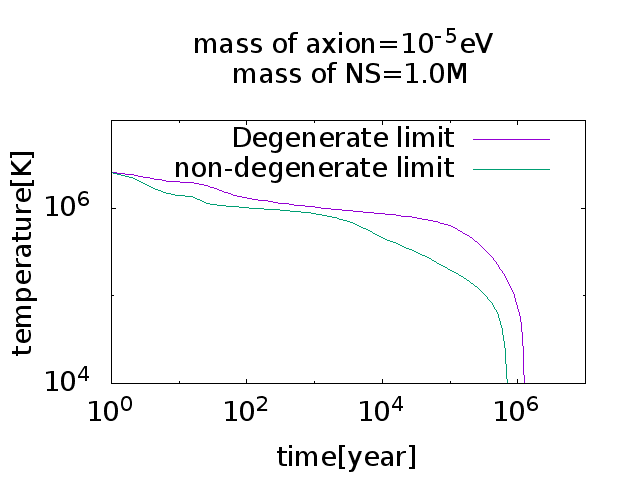}
\includegraphics[width=5.5cm,height=7cm]{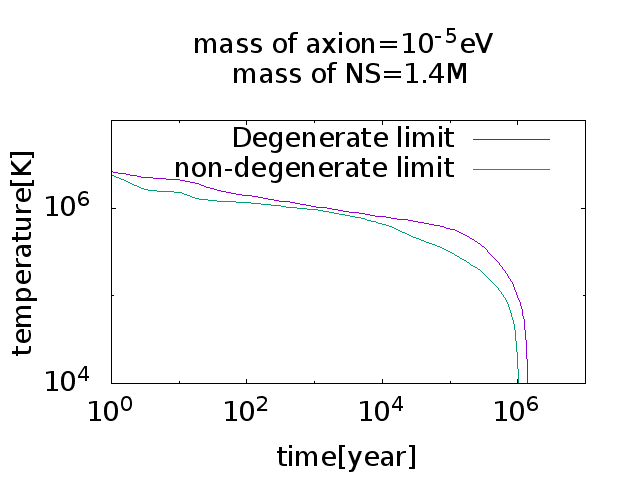}\includegraphics[width=5.5cm,height=7cm]{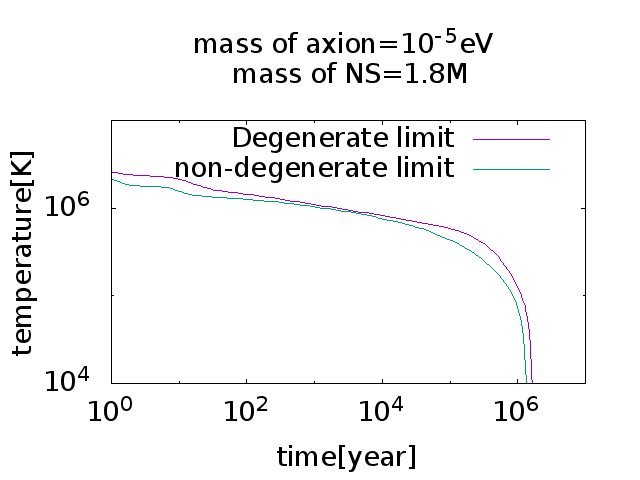}
\caption{Comparison between degenerate and non-degenerate limits for $T$ vs. $t$ plot with $m_{a}$=$10^{-5}\rm{eV}$ and $ \rm{M}=1.0\rm{M}_{\odot}$, $ \rm{M}=1.4\rm{M}_{\odot}$ and $ \rm{M}=1.8\rm{M}_{\odot}$ (from left to right).}
\label{fig:11}
\end{figure}
\begin{figure} 
\includegraphics[width=5.5cm,height=7cm]{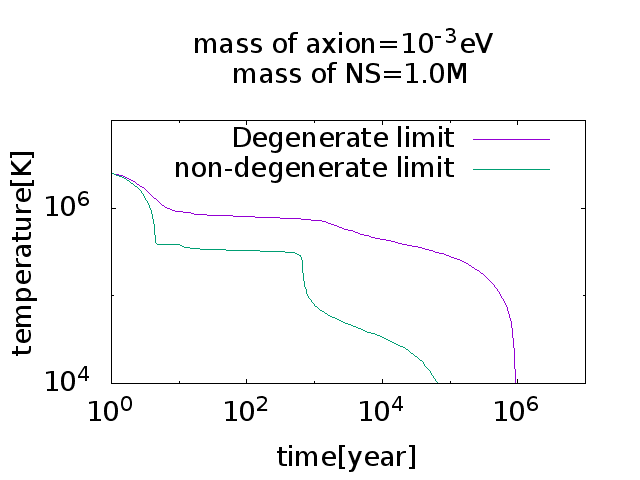}
\includegraphics[width=5.5cm,height=7cm]{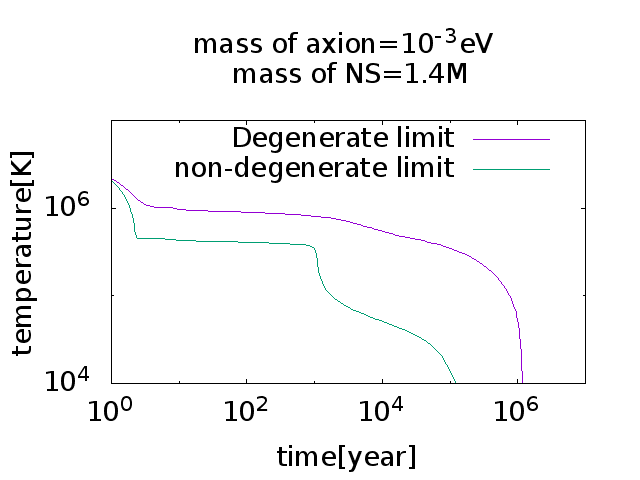}\includegraphics[width=5.5cm,height=7cm]{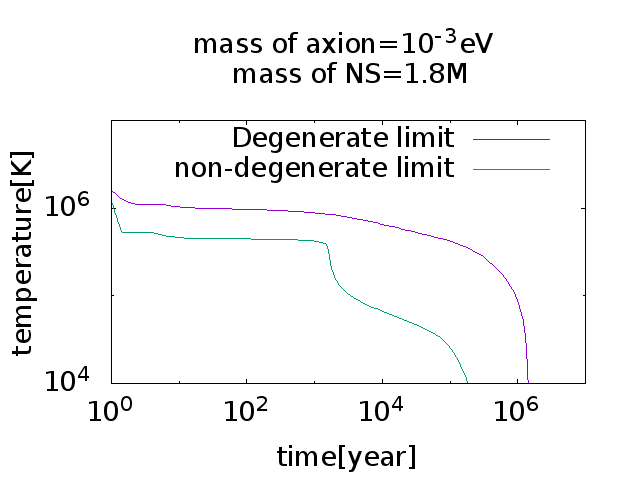}
\caption{Same as in Fig. \ref{fig:11} but for $m_{a}$=$10^{-3}\rm{eV}$.}
\label{fig:12}
\end{figure}

\newpage
\section{Summary and Discussions}
In this work, we have explored the effect of axion emission on the cooling of neutron stars. For the axion emission we consider nucleon-nucleon axion bremsstrahlung process. We have made our analysis for the stars considering degenerate and non-degenerate limits. In earlier such works related to axion cooling of neutron stars, only degenerate process has been considered. For the present purpose, we adopt the APR equation of state. However, one can use other EoS available in the literature. From our analyses, we find that the axion emission effects the neutron star cooling through other estrablished processes such as gamma and neutrino emissions. We have demonstrated this by considering three axion masses, namely $10^{-5}\rm{eV}$, $10^{-3}\rm{eV}$ and $10^{-2}\rm{eV}$ and the neutron star masses of 1.0${M}_{\odot}$, 1.4${M}_{\odot}$ and 1.8${M}_{\odot}$. For comparison with observational results, we used luminosity and temperature of three pulsars namely PSR B0656+14, Geminga and PSR B1055-52. We demonstrate our results by calculating the variations luminosity and temperature of neutron stars with time. While for the neutron star masses 1.0${M}_{\odot}$, 1.4${M}_{\odot}$ the observational points barely agrees with these variations, for the neutron star mass of 1.8${M}_{\odot}$ this agreement appears to improve marginally. From these analyses, we derive a bound on axion masses $m_{a}\leq10^{-3}$ which implies that the decay constant $f_{a}\geq 0.6\times10^{10}\rm{GeV}$. We also find that the cooling patterns for degenerate and non-degenerate cases are different. This difference increases with the mass of the emitted axion. More elaborate studies involving other equation of states are required to obtain better insight of the effect of axion emission on neutron star cooling. This is for posterity.

{}
\end{document}